\documentclass[runningheads]{llncs}
\usepackage{graphicx}
\usepackage{rotating}
\usepackage{multirow}
\usepackage{array}
\newcolumntype{L}[1]{>{\raggedright\let\newline\\\arraybackslash}p{#1}}
\newcolumntype{C}[1]{>{\centering\let\newline\\\arraybackslash}p{#1}}
\newcolumntype{R}[1]{>{\raggedleft\let\newline\\\arraybackslash}p{#1}}

\begin{document}
\title{On Observability and Monitoring of Distributed Systems -- An Industry Interview Study }
\titlerunning{On Observability and Monitoring of Distributed Systems}
%
%
\author{Sina Niedermaier\inst{1} \and
Falko Koetter\inst{2} \and
Andreas Freymann\inst{2} \and Stefan Wagner\inst{1}}
\authorrunning{S. Niedermaier et al.}
%
\institute{Institute of Software Technology, University of Stuttgart, Stuttgart, Germany \email{\{sina.niedermaier,stefan.wagner\}@iste.uni-stuttgart.de} \and
Fraunhofer Institute for Industrial Engineering IAO, Fraunhofer IAO, Stuttgart, Germany \email{\{falko.koetter,andreas.freymann\}@iao.fraunhofer.de}
}
\maketitle              
\begin{abstract}
Business success of companies heavily depends on the availability and performance of their client applications. Due to modern development paradigms such as DevOps and microservice architectural styles, applications are decoupled into services with complex interactions and dependencies. Although these paradigms enable individual development cycles with reduced delivery times, they cause several challenges to manage the services in distributed systems. One major challenge is to observe and monitor such distributed systems.
This paper provides a qualitative study to understand the challenges and good practices in the field of observability and monitoring of distributed systems.
In 28 semi-structured interviews with software professionals we discovered increasing complexity and dynamics in that field. Especially observability becomes an essential prerequisite to ensure stable services and further development of client applications. However, the participants mentioned a discrepancy in the awareness regarding the importance of the topic, both from the management as well as from the developer perspective.
Besides technical challenges, we identified a strong need for an organizational concept including strategy, roles and responsibilities. Our results support practitioners in developing and implementing systematic observability and monitoring for distributed systems.

\end{abstract}
\section{Introduction}

In recent years, many IT departments have successfully migrated their services to cloud computing~\cite{sfondrini2018cloudadoption}. Still, challenges for cloud adoption remain regarding the operation and holistic monitoring of such services~\cite{natu2016holistic}. While conventional IT infrastructure can be monitored with conventional monitoring solutions, cloud environments are more dynamic and complex~\cite{Aceto2013monitoringsurvey}, resulting in a gap~\cite{kinsella2015complexitygap} between the complexity of distributed systems and the capability of monitoring tools to manage that complexity. 

Emerging trends like Internet of Things (IoT) and microservices further increase the complexity, making monitoring a significant barrier for adoption of these technologies~\cite{knoche2019drivers}. 

While newly emerged software tools try to bridge this complexity gap, the way forward for many companies is unclear. We found that there is no research matching new solutions and technologies to different application areas, problems and challenges. While a plethora of new technologies and approaches exists, companies need to be able to relate these technologies to the challenges they face. Processes and good practices are necessary to incorporate new solutions into existing enterprise architectures as well as emerging cloud architectures.

To address this need, we conducted an industry interview study among different stakeholders involved in monitoring, including service managers, DevOps engineers, software providers, and consultants. From the semi-structured interviews, we extracted contemporary challenges, requirements, and solutions.


\section{Related Work}
\label{Related Work}

\begin{figure}[h!]
	\centering
	\includegraphics[width=1.0\textwidth]{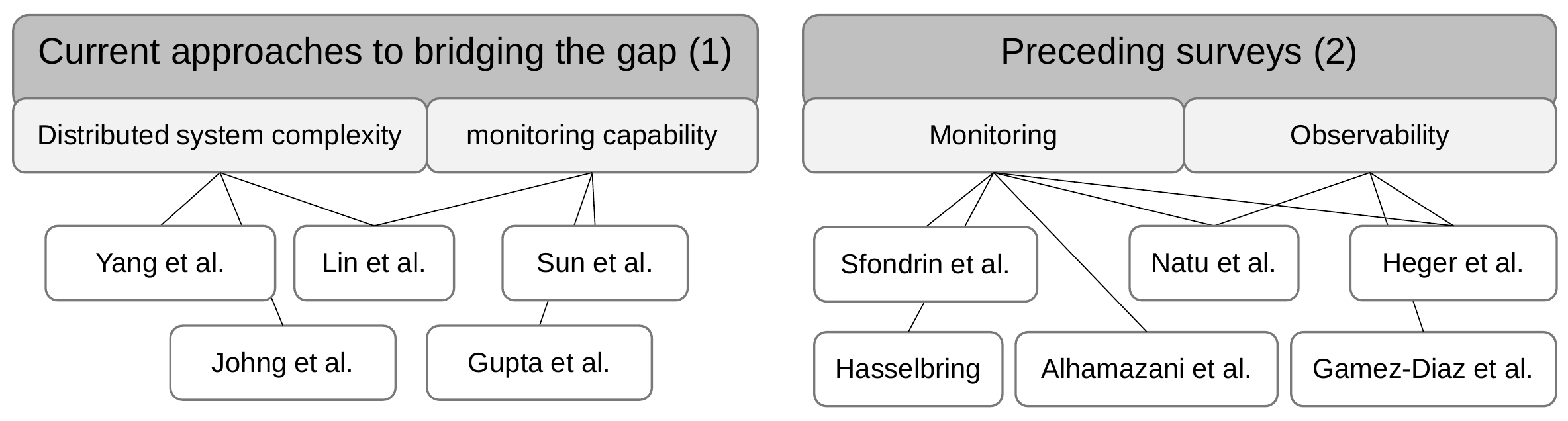}
	\caption{Overview on the related work}
	\label{img:figureRelatedWork}
\end{figure}

To provide context to the survey described in this work, the related work investigates (1) current approaches to bridging the gap between distributed system complexity and monitoring capability as well as (2) preceding surveys regarding monitoring and observability (see Figure \ref{img:figureRelatedWork}).

IEEE defines \emph{monitoring} as the supervising, recording, analyzing or verifying the operation of a system or component~\cite{IEEE1990glossary}. 

The term \emph{Observability} originates in control system theory and measures the degree to which a system's internal state can be determined from its output \cite{gopal1993observability}. In cloud environments, observability indicates to what degree infrastructure and applications and their interactions can be monitored. Outputs used are for example logs, metrics and traces~\cite{picoreti2018observability}.

Yang et al. \cite{yYang2018TransparentlyCapturingExecutionPath} investigate the capturing of service execution paths in distributed systems. While capturing the execution path is challenging, as each request may cross many components of several servers, they introduce a generic end-to-end methodology to capture the entire request. During our interviews we found a need for transparency of execution paths as well as more generally interdependencies between services.

The current trend towards more flexible and modular distributed systems is characterized by using independent services, such as micro- or web services. While systems consisting of web services provide better observability than monolithic systems, services have the potential to enhance their observability and monitoring by giving relevant information about their internal behaviour. Sun et al. \cite{cSun2018ConstraintBasedModelDriven} deal with the challenge that web service definitions do not have any information about their behaviour. They extend the web service definition by adding a behaviour logic description based on a constraint-based model-driven testing approach. During our interviews we identified that the behaviour especially of third-party services needs to be more clearly communicated to assess the impact on service levels and to detect and diagnose faults.

Besides monitoring individual service calls, it is important to predict the runtime performance of distributed systems. Johng et al. \cite{johng2018estimating} show that two techniques, benchmarking and simulation, have shortcomings if they are used separately and introduce and validate a complementary approach. Their approach presents a process which maps benchmark ontologies of simulations. This prove to be inexpensive, fast and reliable. Similarly, Lin et. al \cite{lin2018microscope} propose a novel way of root cause detection in microservice architectures utilizing causal graphs. In our interviews we found that performance is often only known when a system goes live, as the interdependencies between different services and their individual performance are not assessed beforehand. 

Gupta et al. \cite{mGupta2018RuntimeMonitoring} addresses runtime monitoring on continuous deployment in software development as a crucial task, especially in rapidly changing software solutions. While current runtime monitoring approaches of previous and newly deployed versions lack in capturing and monitoring differences at runtime, they present an approach which automatically discovers an execution behaviour model by mining execution logs. Approaches like this that gather information automatically instead of necessitating manual definition are crucial with growing complexity and dynamics of distributed systems.

These works show that for research on closing the complexity gap between cloud environments and their monitoring is ongoing. 
However, these solutions are not yet widely adopted in practice. When adopting new technology to industry application, non-functional requirements such as usability, configurability and adaptability increase in importance. 

In the following preceding surveys in the context of monitoring and observability (2) are described.

Aceto et al. \cite{Aceto2013monitoringsurvey} conducted a comprehensive cloud monitoring survey in 2013, detailing motivations, then-current tool support and open challenges. They identified the need for scalability, robustness and flexibility. The survey correctly predicted the rise in complexity and dynamics of cloud architectures and propose actions to handle these such as root-cause-detection, filtering/summarizing of data, and cross-layer/cross-platform monitoring. Similarly, another early survey of cloud monitoring tools by Fatema et al. \cite{fatema2014toolsurvey} identifies capabilities such as scalability, robustness, interoperability and customizability to find a gap between necessary capabilities and existing tools.

Sfondrin et al. \cite{sfondrini2018cloudadoption} conducted a survey of 62 multinational companies on public cloud adoption. While use of public cloud infrastructure is on the rise, barriers like security, regulatory compliance, and monitoring remain. Regarding monitoring, the survey has shown that half of the companies rely solely on their cloud providers' monitoring dashboard. Participants noted a crucial need for quality of service monitoring integrated with their monitoring tool.

Similarly, Knoche and Hasselbring \cite{knoche2019drivers} conducted a survey of German experts on microservice adoption. Drivers for microservice adoption are scalability, maintainability and development speed. On the other hand, barriers to adoption are mainly operational in nature. Operations department resist microservices due to the change in their tasks. On the technical level, running distributed applications prone to partial failures and monitoring them is a significant challenge.

Gamez-Diaz et al. \cite{diaz2017restfulAPIs} performed an analysis of RestFUL APIs of cloud providers, identifying requirements for API governance and noting a lack of standardization.

While not an empirical study, Natu et al. \cite{natu2016holistic} show monitoring challenges of holistic cloud applications. Scale and complexity of applications is identified as a main challenge. Related to observability, incomplete and inaccurate views of the total system as well as fault localization are other identified challenges. 

Heger et al.\cite{heger2017APM} give an overview of the state-of-the-art in application performance monitoring (APM), describing typical capabilities and available APM software. They found APM to be a solution to monitoring and analyzing cloud environments, but note future challenges in root cause detection, setup effort and interoperability. APM cannot be understood as a purely technical topic anymore but needs to incorporate business and organizational aspects as well.

Alhamazani et al. \cite{alhamazani2015commercialtools} give an insight into commercial cloud monitoring tools, showing state-of-the-art features, identifying shortcomings and, connected with that, future areas of research. Information aggregation across different layers of abstraction, a broad range of measurable metrics and extensibility are seen as critical success factors. Tools were found to be lacking in standardization regarding monitoring processes and metrics.

Comparing preceding surveys regarding monitoring and observability (2) to our work, these surveys focus either on drivers and challenges or on available solutions (in science and commercial tools).
In comparison, our study takes a holistic approach. We provide empirical industry-focused research with in-depth interviews, where we combine different perspectives in order to find out which emerging solutions and strategies are used by companies and to what degree they overcome the existing challenges.
This is necessary to gauge the adoption of new technologies in practice. Moreover, it comprises which challenges these technologies address and which challenges emerge in adoption or are unsolved.


\section{Scope and Research Method}
\label{Scope and Research Method}

\textbf{Study Design:}
To structure our research, we applied the five-step case study research process as described by Runeson and Hoest \cite{Runeson2008}. Our research objective can be defined as follows: \textit{Analysis of the contemporary challenges of monitoring and operating distributed systems for the purpose of deducting requirements and mapping existing solutions and strategies from the viewpoint of different stakeholders of monitoring systems and tool providers.} Table \ref{tab:tableResearchQuestion} summarizes the research questions:\\

\begin{table}[h!]
\scriptsize
  \begin{center}
    \caption{Overview of the research questions}
    \label{tab:tableResearchQuestion}
    \begin{tabular}{  m{1cm}  m{8cm}  }
      \hline
      \textbf{RQ1} & Which contemporary challenges exist in monitoring distributed systems? \\
      \hline
      \textbf{RQ2} & Which requirements do stakeholders have for a monitoring and observability concept for distributed systems? \\
      \hline
      \textbf{RQ3} & What are technical and organizational strategies and solutions in companies? \\
      \hline
    \end{tabular}
  \end{center}
\end{table}

To answer our research questions, we applied the qualitative method of semi-structured interviews. They allow us to explore the individual challenges stated by the participants and to analyze the underlying relations by providing a basic agenda. At the same time, interviews enable dynamic interaction based on the background of our experts and their responses~\cite{singer2008software}.\\
In total, we conducted 28 semi-structured interviews of 45 minutes on average. The interviews were completed between February and April 2019. To achieve a balanced distribution of interviewees, first, we considered users using monitoring solutions and tool providers offering monitoring solutions (see Table~\ref{OverviewParticipantsCompanies}). Second, we ensured that solution providers and users are related to different domains and focus to get diverse perspectives of monitoring solutions. Apart from the tool providers, we covered further domains such as software and IT service, IoT, telecommunication, insurance, and IT consulting. The users have been selected from different points of view in the application stack and different roles like DevOps and support engineers along with product owners and managers. The recruiting of participants was achieved by personal industry contacts as well as by acquisition on developer conferences.

\begin{table}[t]
\caption{Overview about participants and companies }\label{OverviewParticipantsCompanies}
\scriptsize
\begin{tabular}{
    C{0.6cm}
    L{1.8cm}
    C{1.2cm}
    C{0.6cm}
    L{3.4cm}
    L{3.9cm}
}

\hline
\textbf{CID} &  \textbf{Domain} & \textbf{Staff} & \textbf{EID} & \textbf{Expert role} & \textbf{Focus}\\
\hline

\textbf{C1} &  IoT &\(>\) 100T & E1 & Product Owner & APM Solution\\
&&& E2 & Lead Architect & Cloud Infrastructure\\
&&& E3 & Product Owner & Connectivity Backend\\
&&& E4 & Service Manager & Support and Operation of IoT Solution\\
&&&E5&Cloud Architect&IoT Backend\\
\hline
\textbf{C2}&IoT&100-1T&E6&IoT Consultant/Architect &Consulting of IoT Projects\\
&&&E7&Open Source Developer&Cloud Service \\
&&&E8&DevOps Engineer&Cloud Service\\
&&&E9&DevOps Architect&Cloud Service\\
&&&E10&Product Owner &Cloud Service\\
&&&E11&Project Lead&IoT Project\\
\hline
\textbf{C3}&IoT&10T-100T &E12&Manager & IoT Platform\\
\hline
\textbf{C4}&IoT&10T-100T &E13&IoT Solution Owner &Cloud Service\\
\hline
\textbf{C5}&Telecom.&10T-100T &E14&Product Owner &Monitoring Platform\\
\hline
\textbf{C6}&Software and IT Services&\(>\) 100T &E15&Former Chief Technology Officer& Software Development and Operations Tool\\
&&&E16&Technical Lead IT-Operations& Operation Solution and Event Management\\
\hline
\textbf{C7}&Applied Research&100-1T &E17&DevOps Engineer&Insurance Service\\
& & & E18&Developer&Front- and Backend of Fleet Management\\
\hline
\textbf{C8}&Tool Provider&1T-10T &E19&Sales Engineer&APM Solution\\
\hline
\textbf{C9}&Tool Provider &10T-100T &E20&Strategic Officer&Infrastructure Monitoring Tool\\
&&&E21&Support&Infrastructure Monitoring Tool\\
\hline
\textbf{C10}&Tool Provider&100-1T &E22&Developer and Architect&Infrastructure Monitoring Tool\\
\hline
\textbf{C11}&Tool Provider&100-1T &E23&Developer&Monitoring Tool\\
\hline
\textbf{C12}&IT Service Insurance&1T-10T &E24&Divisional Director Monitoring&Performance-Monitoring\\
\hline
\textbf{C13}&IT Consulting&1-25&E25&Developer and Architect&IT Consulting Monitoring\\
\hline
\textbf{C14}&IT Consulting&100-1T&E26&Chief Executive Officer&Business Process Monitoring\\
\hline
\textbf{C15}&Software and IT Services &10T-100T&E27&Solution Architect&Open Source Technology Provider\\
\hline
\textbf{C16}&Software and IT Services&10T-100T&E28&Developer&Cross-Stack Instrumentation for Monitoring and Debugging\\
\hline
\multicolumn{6}{c}{*CID = Company ID, *EID = Expert ID}
\end{tabular}
\end{table}

\textbf{Preparation for Data Collection:}
To conduct the semi-structured interviews, we created an interview guide~\cite{niedermaier2019guideline}. The guide is structured in different thematic blocks to group the individual questions. 
The interviewees were pre-informed about scope and procedure of the interviews. Besides the information to treat their transcripts as confidential, we asked to record the interviews to create transcripts if permitted. Moreover we informed them about the possibility to review their transcript to assent to the information given in the interview. 

\textbf{Data Collection:}
From the 28 interviews, 15 were conducted 'face to face' and 13 via remote communication. The interviews were held in German, except for two interviews in English. While 21 interviews have been audio recorded, for the remaining interviews two researchers created protocols to reduce researcher bias. During the interviews, we loosely followed the interview guide accordingly to the answers and to the participants' focuses. After manually transcribing the interviews, we sent the transcripts to the participants for review, where they had the possibility to correct unintended statements or remove sensitive data.

\textbf{Data Analysis:}
For the analysis of the individual transcripts, we encoded the material to extract important categories regarding our research goal. For this purpose, we followed Mayring's approach of qualitative content analysis~\cite{mayring2014qualitative}. We openly encoded the transcripts by applying inductive category development, where we analyzed the transcripts on sentence level. Usually, one code was assigned to different sentences in a transcript and furthermore one sentence could be assigned to more than one code. During analysis we formed hierarchies of codes and sub-codes. In several iterations, the codes were revised, split or merged.

\section{Results and Discussion}
\label{Results and Discussion}
This section presents the aggregated findings from the interview analysis with the focus on our research questions defined in Section \ref{Scope and Research Method}. We created a hierarchy of categories as an abstraction of the codes defined during the analysis of the transcripts. This paper presents the top-level hierarchy of the identified challenges, requirements, and solutions. In the following, we describe the different codes generated according to the research questions and illustrate the answers given for the codes with some exemplary statements from the experts~(see EID Table~\ref{OverviewParticipantsCompanies}).

\subsection{Challenges}
\label{Challenges}
The first research question (RQ1) aims to understand the challenges our participants deal with in the field of distributed systems and which implications are further related with these challenges. We identified a set of nine challenges (Cx) and their corresponding implications which are described in the following.

\textbf{Increasing dynamics and complexity (C1):}
The emerging trend of microservice architectures, cloud deployments, and DevOps increase the complexity of distributed systems. While the individual complexity of a microservice is reduced, the complexity of the interdependencies of microservices and the dynamic components within a distributed system cause more operational effort. This dynamic environment is not manageable manually and traditional approaches such as Configuration Management Database (CMDB) \cite{colville2006cmdb} are not sufficient anymore:
\textit{"CMDB are often based on polling and get the state of the system once a week. In one week, a lot has happened in the cloud system, which a CMDB can not cover."}~(E16). This issue does not only include cloud native microservice architectures but also historically grown systems, where an overview of service the dependencies is missing. In addition, some participants stated an underestimation of the dynamic complexity of their systems. This caused that in case of a problem (especially for the diagnosis of context dependent or non-permanent faults) the average duration for detection and recovery took too long.

\textbf{Heterogeneity (C2):}
Todays distributed systems consist of several layers: from  application to infrastructure technologies like containers, VMs or even serverless environments. These layers are developed and operated by heterogeneous teams. Moreover, as stated by our participants, systems often contain legacy and modern service technology in parallel, where additional tooling is necessary to integrate legacy components. With regard to multi-tenant systems, some participants experienced a noisy-neighbour-effect, where one tenant monopolizes resources and negatively affect other tenants on the same infrastructure. However, in this case the participants were not able to separate views among different tenants. In terms of technological heterogeneity and speed of innovation, the participants had divisive opinions. For distributed systems developer can choose the most suitable technology on the one hand, but on the other hand, the technological heterogeneity complicates the consistent application of monitoring tools. Furthermore, other participants have criticized the speed of innovation and some require a slow-down of technology hypes by defining regulations. The heterogeneity in these different areas is leading to a missing overview of the overall system, it´s individual components and the requests processed. 

\textbf{Company culture and mindset (C3):}
Most of the participants believe that culture and mindset aspects referring to monitoring are essential. Several even stated that this aspect is more challenging than technical aspects. Furthermore, some interviewees also mentioned that a holistic transparency to apply monitoring is often not intended. This gives for instance rise to danger of being blamed in retrospect for a failure. Often, the participants described that teams do not have an overview outside of their own area, for example of the business context of their service. This caused isolated monitoring and operation concepts without context to customer solutions and related requirements. Overall, collaboration and communication between teams and the perspective from which they develop and operate their services are often weakly pronounced. This illustrates the following statement (E22): \textit{"It is usually not the ignorance or the inability of people in the company, but the wrong point of view. Often the developers are so buried in their problem environment, so engrossed in their daily tasks that they can no longer afford to change themselves."}

\textbf{Lack of central point of view (C4):}
Participants stated limited possibilities in terms of visibility and dependencies to other services and teams. This results in turn in a missing system-wide overview. E6 describes for instance such situation: \textit{"If it comes to problems such that there are many user complains because the system is not working properly, everyone went for troubleshooting. Due to the lack of an overview, it was difficult to diagnose the faults. It took several escalation rounds and teleconferences to discuss where the fault is located."} At the same time, we identified a lack of a responsible persons in charge to generate overall views and thus to enable individual teams to collaborate. Another point mentioned is the missing transparency about the impact on availability and performance of integrated components from 3rd parties which are often part of a distributed system. Due to the fact that for such components service parameters are usually not accessible, blind spots remain and prevent an overall monitoring.

\textbf{Flood of data (C5):}
Participants mentioned the overwhelming flood of data coming from the distributed system, which is constantly in change. The identified challenge is to create meaningful conclusions from customer alerts and how to prioritize them. This shows the following statement (E17):\textit{"The volume and amount of alerts are currently challenging, we are not able to prioritize the customer impacting ones."} Moreover, for problems, where one request has to be handled by multiple components that are developed by independent teams, it is very complex to identify the location of faults, including the responsibilities to fix it. In more detail, many participants described the complexity in correlation of metrics and timestamped logs from multiple services which is often accompanied by insufficient metadata. In addition, participants stated an absence of a comprehensible dashboard that enables navigable views through the data.

\textbf{Dependency on experts (C6):}
The process of fault detection and diagnosis, which are often manually performed, seems to be highly dependent on knowledge of individual experts about design and behaviour of the systems. As the following statement from E11 illustrates, these experts appear as a 'source to debug': \textit{"This form of troubleshooting depends highly on the expert knowledge of the team members} [...]\textit{. Mostly, the knowledge about the structure of the service is currently more crucial than a monitoring which specifically indicates 'search at this point'."} This challenge again outlines the missing systematic development of monitoring systems in supporting humans in fault detection and diagnosis.

\textbf{Lack of experience, time and resources (C7):}
Many participants described the challenge of mastering microservice technologies and the DevOps paradigm, which require additional effort for operation, but at the same time, skilled DevOps engineers are missing. Especially, the short time to market results in a prioritization of features and in a disregard of non-functional requirements like availability or performance. Most participants mentioned the limited time as reason for an iterative, often reactive development of system observability and monitoring.

\textbf{Unclear non-functional requirements (C8):}
According to the interviewees, non-functional requirements like availability and performance, also referred to as Quality of Service (QoS) or Service Levels (SL), are often not or insufficiently defined and controlled. In addition, some participants commented that teams are often not aware of their major QoSs as well as of their importance in the context of what needs to be measured and monitored. The related reasons for that is due to missing or unprecise customer requirements or due to the lack of awareness regarding the importance of non-functional requirements (E6):\textit{ "It is very important to think about service levels or KPls and to define them in a certain way. This is often underestimated. In many projects it can be determined that the project managers only have a purely technical view of the system without being aware of the availability and performance that is needed."} Another reason we identified is the complexity to define overall availability and performance goals, which then have to be converted into goals for component services. This is further intensified by time constraints that lead to reactive implementations as stated in the following statement (E6):\textit{"Many development teams are under pressure to bring the service to market as quickly as possible. So the teams usually start developing without specific customer requirements and end up in production without any systematically derived requirements."} The lack of requirement definitions leads in turn to missing feedback loops (E4): \textit{"}[...]~\textit{where the quality control in the service provision is missing". }

\textbf{Reactive implementation (C9):}
As stated in several interviews, the unclear requirements and the lack of sufficient indication and control often leads to failure. In these cases, the development of monitoring was triggered by an failure in production, where the teams recognized a lack of observability to diagnose customer failure or even to detect them. In fact, customers often received inadequate service levels. In several examples, the teams were occupied only with troubleshooting, which in turn resulted into ad-hoc solutions, instead of creating systematically derived monitoring solutions. Moreover, we identified that teams run into same problems, where labor-intensive development of monitoring for individual services are created and synergy effects of sharing knowledge, expertise and good practices are not used. A further reason for reactive implementation is that during development, the developers did not have enough knowledge about the complex interactions in production and therefore blind spots remained until operation.

\subsection{Requirements and Solutions}
\label{Requirements and Solutions}
\begin{figure}
	\centering
	\includegraphics[trim={37pt 20pt 0pt 20pt}, width=373pt]{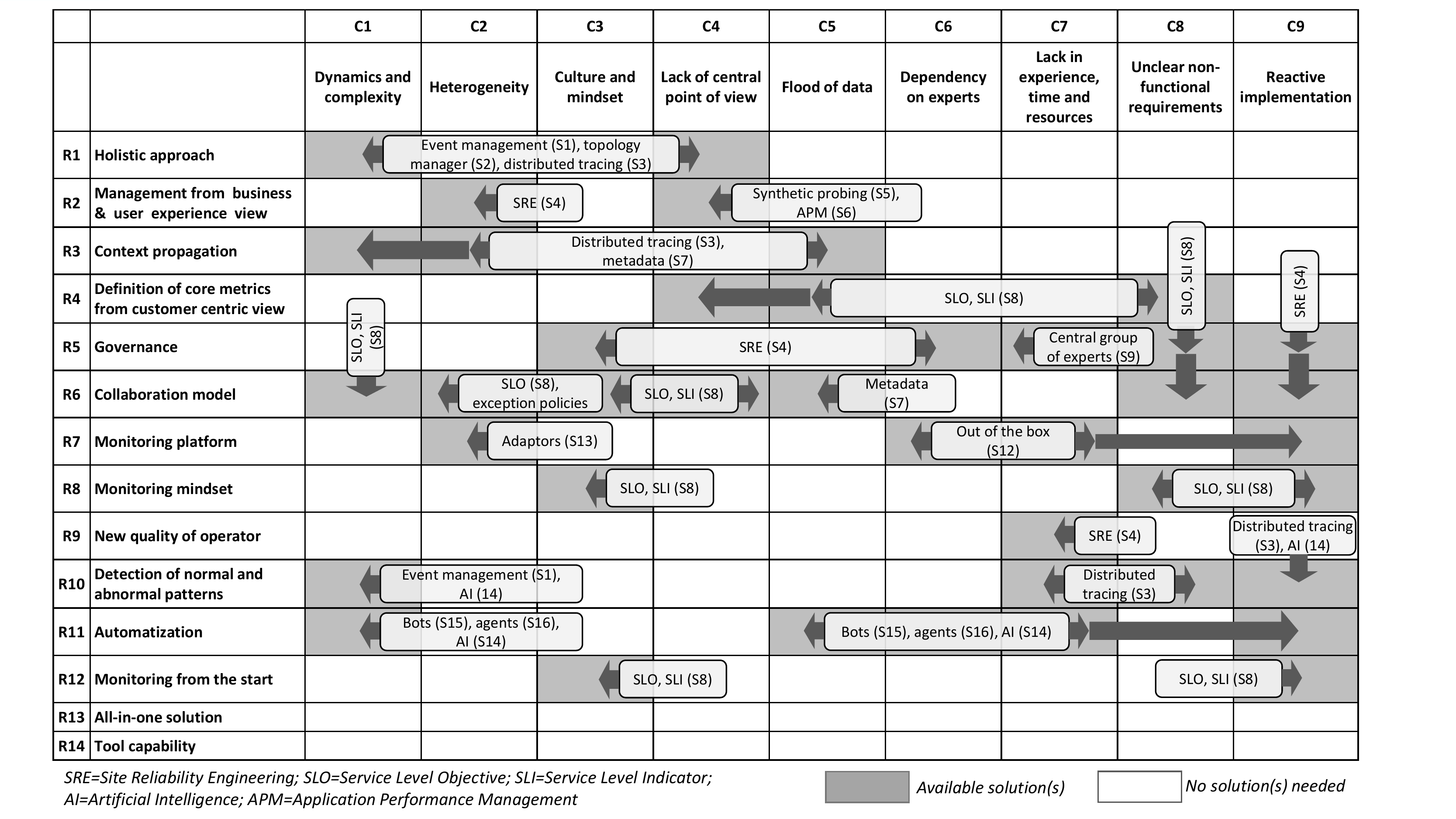}
	\caption{Overview of challenges, requirements and solutions}
	\label{img:tableChallengesRequirementsSolutions}
\end{figure}

Regarding RQ 2 and RQ 3, the following requirements (Rx) and possible solutions (Sx) are listed. Rx and Sx have been extracted from the interviews. The mapping of the previous described challenges, the corresponding requirements as well as the solutions are shown in Figure~\ref{img:tableChallengesRequirementsSolutions}.

\textbf{Holistic approach (R1):} Along with C4, the interview participants stated that they characterize monitoring \textit{"}[...] \textit{as holistic problem and try to come up with a holistic approach to ensure observability} [...]\textit{"}~(E28). Therefore, it is necessary to enable collaboration and communication along different system layers and teams. We worked out that a common and central view is required which assists the implementation of a system-wide diagnosis and fault detection. One solution stated by interviewees is an event management system (S1), also referred as 'manager of managers', that enables an overall view of the system state. This allows to correlate events for event reduction. Other solutions mentioned are topology managers and architecture discovery modeling (S2). These enable to dynamically map transactions to underlying infrastructure components. Moreover, distributed tracing (S3) was emphasized as solution, which records the execution path of a request at runtime by propagating request IDs \cite{sambasivan2016principled}. This solution enables to capture causal relationship among events on the execution path. It allows to create a \textit{"}[...] \textit{bird’s eye view, to find out what is going on with a user request} [...]\textit{"}~(E28). Distributed tracing can also be applied for diagnosis of 3rd party components where source code is not available. Hence, the participants mentioned that the instrumentation in order to propagate trace IDs through individual services, developed by different teams, is at the moment not consistently assured.

\textbf{Management from business and user experience view (R2):}
Several participants described the trend moving from isolated monitoring of individual services to a context dependent view from the perspective of a customer or business application. In addition, some stated to apply Google´s approach of Site Reliability Engineering (SRE) (S4) (E28):
\textit{"With our SRE approach in mind, we care about the user experience and these are the golden paths we want to improve. I do not necessarily care about what is going on underneath, as soon as the user is not experiencing any errors, latency or unlikely indicators."} SRE \cite{beyer2016site} takes aspects of software engineering and applies them to infrastructure and operations problems with a focus on customer experience. Moreover, several interviewees stated to perform synthetic probing. This is known as end-to-end monitoring (S5) that enables emulation of real user behaviour to measure and compare its availability and performance. In general, many participants referred to apply Application Performance Management (APM)~(S6) which comprises methods, techniques, and tools to continuously monitor the state of a system from an application-centric view. The APM allows diagnosing and resolving especially performance-related problems. 

\textbf{Context propagation (R3):} To provide a holistic approach and to be able to detect customer impacting events context propagation is needed. The system propagates relevant context in form of metadata, such as IDs or tags along the execution flow of a request through services. Alongside with distributed tracing (S3), adding metadata (S7) to metrics and logs are further examples. This helps to localize the underlying fault in the flood of data generated by the systems by providing contextual information (E28):\textit{"With context propagation, you can easily point to the root cause of the issues. For example, there is some additional latency and you can see that this other particular database call is causing the additional latency. You can automatically inspect and ping the right time and component}~[...]\textit{"}.

\textbf{Definition of core metrics from customer centric view (R4):}
A way to systematically define metrics for individual services is the concept of establishing service level metrics. Some participants apply Service Level Objectives (SLO) and Service Level Indicators (SLI) (S8) as part of a SRE approach. While SLOs describe business objectives by defining the acceptable downtime of a service from a user perspective, the SLI in turn enables to tie back metrics to the business objectives. The interviewees express an essential need for a systematic definition of these metrics, but at the same time they struggle with their implementation (E28): \textit{"It is a lot of work to figure out the right SLO, which is a very long process. Not everybody is interested in this.} [...] \textit{It is hard to introduce this concept at a later time. This is  creating tension in teams, because they are saying: "This is not what you have promised us". But the problem is, if nobody actually formulated what the promise was."}

\textbf{Governance (R5):} To foster the previous requirements, several participants outlined the need for a governance that defines a strategy including roles, responsibilities, processes and technologies for monitoring and observability. This should comprise clear formulations of a minimal set of indicators that have to be monitored from every service. A further requirement is to claim observability of a service as an acceptance criteria for development and operation. The participants mentioned that developing and applying governance needs several iterations and has to be continuously adapted in terms of the company strategy. Especially for services running in the cloud, guidelines have to be defined because \textit{"Cloud is standardization"}~(E16). Concurrently some participants criticized the introduction of tooling standards and a slow-down of development by oversized governance regulations. Some companies already have an own department and group of specialists with the central responsibility of monitoring~(S9). To provide a strategy, participants mentioned to align their governance to SRE principles and guidelines (S4), which in some examples already evolved into a self-regulating system. A commonly mentioned issue addresses the creation of community of a practice~(S10) to share good practices and lessons learned.

\textbf{Collaboration model (R6):}
Some participants described a collaboration model as an essential base for communication and efficient diagnosis processes along different teams. As part of that, exception policies and taxonomies for anomalies (S11) need to be defined. To efficiently work together during diagnosis metadata (S7), capturing the causal relationships and providing context are needed. Moreover, some participants stated that a (E28): \textit{"}[...] \textit{common language, called SLO and SLI"}~(S8) is base for their team collaboration.

\textbf{Monitoring platform (R7)}: Several interviewees required a unified monitoring platform to increase operational efficiency. This includes out of the box (S12) and standardized components which can be used modularly and are customizable to specific needs of the individual services. Monitoring and its default setup also needs to enable the \textit{"democratization of data"} (E19), for example to offer a standard API to deploy adaptors (S13) for different technologies. This default can therefore
\textit{"}[...] \textit{create a kind of governance that is not strict.}"(E20).

\textbf{Monitoring mindset (R8):} The prerequisite is to increase the importance of observability and monitoring of distributed systems. Without an increasing awareness, isolated ad-hoc solutions will remain, which do not enable sufficient service provisioning and diagnostics. One mentioned solution to increase the importance of non-functional attributes, like availability and performance, can be reached by setting and controlling SLOs (S8). Thereby, stability and feature development can be controlled from management perspective. Further, the participants outlined a need to equalize functional and non-functional requirements.

\textbf{New quality of operator (R9):} Many participants mentioned a lack of academic education of the operators. Accordingly, operators need to increase their skills especially of being able to cope with automation tasks. Achieving this, the company has to increase the awareness of monitoring and to promote proper responsible operators. Some participants highlighted the need for Site Reliability Engineers (SRE) (S4), who are able to work on operation and infrastructure tasks as well as software engineering aspects.

\textbf{Detection of normal and abnormal patterns (R10):} Nearly all interviewees spoke about anomaly detection as an important task for monitoring to differ between what is normal and abnormal behaviour. Different solutions are mentioned, such as event management (S1), to correlate events from different parts of the system. For correlating events, predictive analytics and artificial intelligence (AI) (S14) are in use. Some participants discussed the problem of differentiating normal behaviour of a service. In this context, some participants considered distributed tracing (S3) to indicate performance measures and iteratively develop guarantees for their services by setting service levels. The most advanced method for anomaly detection is AI (S14). Almost all participants appreciated its enormous potential to master the complexity and the flood of data generated by distributed systems. However, many interviewees pointed out that sufficient preconditions for the use of AI are still missing in practice. Primarily, the right data has to be collected, the quality of data has to be ensured, the context needs to be propagated and data has to be stored centrally. Concerns in terms of the cost value ratio of AI approaches and their reliability remain (E28): \textit{"I don´t know if we will ever have a solution that we can rely on confidently"}.

\textbf{Automation (R11):} The increasing dynamics and complexity within distributed systems, caused by the upcoming microservice architecture and shorter lifecycles of components, is not manually controllable as a whole any more. Therefore, automation is indispensable to observe and monitor a distributed system. Especially recurring problems can be automatically solved and basic monitoring techniques can be automatically implemented. Bots (S15) and agents (S16) can realize automation, as they act by themselves. In combination with AI (S14), bots and agents can be more efficient and more precise in their tasks such as collection and analysis of traffic data.

\textbf{Monitoring from the start (R12):} Monitoring is a prerequisite for any development and operation. Many interviewees indicated to consider monitoring from the start. It should be the part of any design. Some participants quote to integrate SLOs and SLIs (S8) \textit{"}[...] \textit{in the design time. As soon as there is a new service you have a section in the design doc., where you can see these are the promises, they may change over time to reduce toil. We start the conversation very early on."}(E28). This might enhances the awareness for monitoring and can change the company culture towards a monitoring mindset. While IT departments should see monitoring as an integrated part, the management needs to be the key driver to implement such a mindset.

\textbf{All-in-one solution (R13):} Solutions covering all monitoring functionalities in one solution were mentioned in the interviews. However, the reality shows that such solutions do not exist. In the best case, the market offers solutions which provide basic functionalities for monitoring such as performance measurement or logging. In more detail, they often provide the capability to easily expand the solutions, for instance by combining and integrating other software solutions. This can also comprise new standards, technologies and other already existing solutions. Hence, all-in-one solutions represent in this context a combination of several solutions and technologies. Nevertheless, to realize such an encompassing solution, it needs to avoid or substitute isolated solutions with standard monitoring software, open standards and modern technologies to reach an encompassing solution in the future.

\textbf{Tool Capabilities (R14):} In this paragraph, we summarized different tool capabilities, mainly non-functional requirements, mentioned by the participants. An often stated requirement is real-time monitoring, where changes and impacts are being directly monitored without delay. Therefore, the necessary information can be provided for appropriate response (e.g. real-time alerts to reduce reaction time). A further requirement addresses the use of open standards (e.g. JSON or standard monitoring functionalities), which is motivated by being adaptable and flexible due to new technologies and standards. This also fosters the maintainability and portability of monitoring solutions by being easily transferable to other distributed systems. Associated with that, scalability is of a particular importance to cope with large and dynamic distributed systems. While the management of the dynamics within a distributed system needs to be addressed, reliability and availability of the monitoring is highly demanded. For example, health functionalities, such as the current status of the system, needs to be available all the time. Moreover, tools need to support multi-tenant management. This requirement specifically addresses the ability of tenant specific views and individual permission management. A further mentioned aspect is the importance of the security of the monitoring tools itself. The more agents are used and the higher the integration depth is, the more 'backdoors' might be open and the higher the possible negative impact could be. Thus, security aspects such as prevention actions need to be realized. A minimally invasive approach needs to be followed, where the changes in an existing system are limited. This might be opening just a minimum of relevant ports. In addition, a careless deployment and configuration of monitoring agents have been mentioned as potential problems which might causes instability and an increasing network load. 

\section{Threats to Validity}
\label{Threats to Validity}
For \textbf{internal validity}, there is a risk that the participants did not state the true situation or their opinion. However, this risk is rather small, because we were ensuring the anonymity of the interviews and the participants seemed not to be worried to talk about negative aspects of their product or company.
Another threat to internal validity are potentially misunderstood concepts used within the questions. Therefore, we provided additional explanations for important concepts. Otherwise, we asked questions to clarify terms used by the participants that could have a domain or company specific meaning. To reduce researcher bias and therefore to increase the interpretation validity, every transcript was reviewed by at least one additional researcher. Furthermore, our participants had the chance (and took it) to adjust statements in their transcript that were incorrect, indistinct or contained sensitive data.

To increase \textbf{external validity}, we asked participants not exclusively based in Germany but also participants coming form international companies with diversity in terms of domain and size. Additionally, with our participants we are covering different roles, coming from different layers of the application stack as well as including providers of monitoring solutions and consultants advising companies and teams in integrating monitoring solutions. Therefore, it was possible to generate an overall view of the complex relations in terms of technical and organizational aspects leading to challenges as well as requirements and solutions. Still, as we performed qualitative research, we do not claim our results to be generalizable. 

\section{Conclusion}
\label{Conclusion}

Our research objective was to explore challenges, requirements and contemporary good practices as well as solutions in terms of monitoring and observability of distributed systems. Therefore, we conducted interviews with 28 software professionals from 16 organizations.
We identified that monitoring and the observability of distributed systems is not purely a technical issue anymore but becomes a more cross-cutting and strategic topic, critical to the success of a company which offers services. Development and deployment paradigms of microservices, DevOps and cloud are creating maximal independence and specialization resulting in isolated monitoring and observability solutions, not allowing to manage a service from a customer or business centric view. Most companies have already solutions and good practices in place, but in many cases they remain isolated approaches due to siloed company structures.
With reference to the findings of the contemporary state of practice, we see a need for further work on good practices and real world-examples for aligning business goals with technical metrics to break down silos and enable efficient development and operation. Furthermore, researchers can take these results into account for designing industry-focused methods.



%
%
%
%
\bibliographystyle{splncs04}
\bibliography{bibliography.bib}

\end{document}